# Tailoring Mie Resonances in Cupric Oxide Particles for Use as Nanoantennas

*Sundaram Bhardwaj Ramakrishnan*[⊥]*, Ravi Teja Addanki Tirumala*[⊥]*, Farshid Mohammadparast*[⊥]*, Swetha M. Arumugam*[§]*, Marimuthu Andiappan\**[⊥]

**Affiliations:**

[⊥] School of Chemical Engineering, Oklahoma State University, Stillwater, OK, USA.

[§] Department of Chemistry, PSG College of Arts & Science, Coimbatore, Tamil Nadu, India.

\* **Corresponding Author**, Email: mari.andiappan@okstate.edu

## ABSTRACT

The field of nano-optics has grown with plasmonic metals. Metals such as silver, gold, and copper nanoparticles, can concentrate electromagnetic (EM) fields at the nanoscale, due to the special property called localized surface plasmon resonance (LSPR). This laid the foundation for a wide range of applications, including nanoscale optics, solar energy harvesting, photocatalysis, and biosensing. However, there are inherent problems associated with plasmonic metals, such as high heating losses, and their inability to be scaled-up like semiconductor fabrication processes. In addition, the field enhancement is restricted only to electric fields. All together these inhibit the broader use of PMNs in practical applications. In this work, we report submicron cupric oxide (CuO) particles with a medium refractive index that can exhibit strong electric and magnetic Mie resonances with strong extinction/scattering cross-sections comparable to or slightly exceeding those of their plasmonic counterparts. Through the development of particle synthesis techniques with strong shape and size control, optical spectroscopy, and finite-difference-time-domain simulations we show that the Mie resonance peak wavelengths are size- and shape-dependent. This gives tunability in the visible to near-infrared regions for harvesting a wider fraction of the solar spectrum. Therefore, submicron CuO particles exhibit strong potential in emerging as high-performance alternatives to PMNs. The strong electric and magnetic Mie-resonance-mediated nanoantenna effect attribute that CuO particles can be potentially used in a plethora of applications, including surface-enhance Raman spectroscopy, metamaterials, photocatalysis, and photovoltaics.

# INTRODUCTION

The idea of optical nanoantennas is an evolving concept in nano-optics. Similar to radio waves and microwave antennas, their objective is to localize EM energy by transforming free propagating radiation, and vice versa. Optical nanoantennas lay the foundation to exploit the unique properties of metal/semiconductor nanostructures. These materials have a strong ability to couple surface plasmons at optical frequencies. Research in nano-optics and plasmonic materials has generated considerable interest in optical antennas. The advancement in optical antennas is limited due to their small scale. Antennas can enhance and facilitate several photophysical processes (Figure 1). In light-emitting devices (LED) combination of charge carriers (electron-hole) pairs can generate a photon (a). In photovoltaics, the reverse of LED, in which incoming light results in charge separation (b). From both cases, we can see that optical antenna can couple with incoming light and localize electric-field make the process of charge transfer efficiency. In addition, its scope can be extended to spectroscopy. The incoming EM field polarizes the molecule of interest, which generates a response through radiation. This emitted radiation can serve as a method for chemical forensic identification as it is related to the material's energy structure (c). In the third case, the antenna facilitates both emission and excitation efficiently. Optical antennas usually are employed to increase the absorption cross-section and quantum yields in photovoltaics, boost efficiency in photochemical and photophysical processes. Since we are interested in nano-scale in this work, we need to develop optical nanoantennas from materials since they differ due to shape, size, and resonant properties.

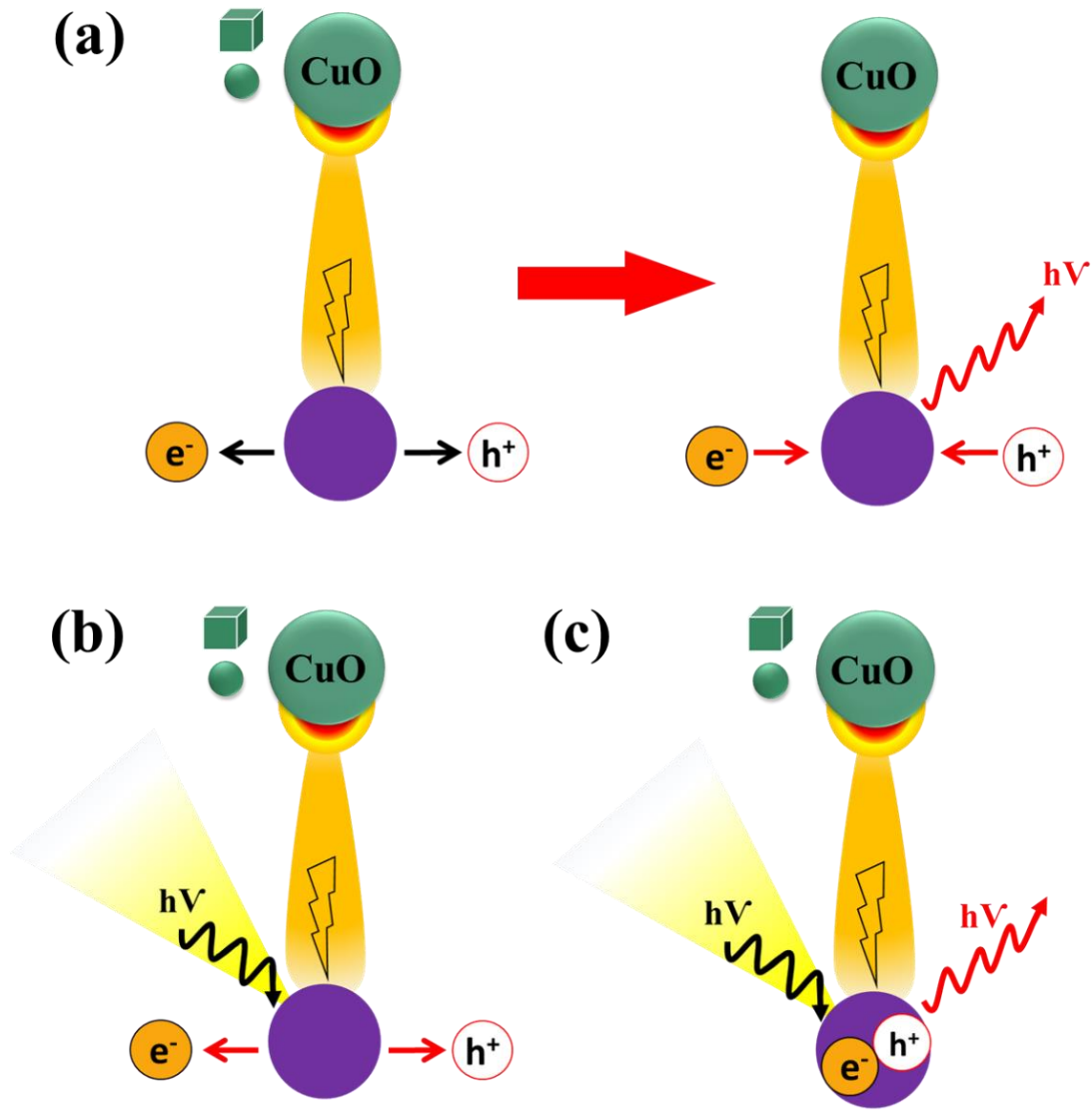

**Figure 1:** Antenna-based optical interactions: (a) antenna-coupled LED, (b) antenna coupled photovoltaics, (c) antenna-coupled spectroscopy.

To develop nanoantennas, people have relied on plasmonic metal nanostructures (PMNs). These materials exhibit a high extinction cross-section due to localized surface plasmon resonance (LSPR). Since the LSPR is sensitive to geometry (shape and size) and physical environment. This property makes them suitable for applications like sensors,[1,2] nano-, and micro-optical devices,[3] photocatalysis[4–10], and photovoltaics.[11–14] The PMNs, by nature, have high extinction cross-sections with resonances in the visible and near-IR, can act as nanoantenna's via EM field enhancement effect that has promising applications in third-generation photovoltaics namely organic, quantum dot, dye-sensitized, and perovskite-based solar cells.[11,12] However, PMNs (Ag and Au) are scarce, expensive, and pose limitations such as high heating losses, and their inability to be scaled-up similar to semiconductor fabrication.[15] To overcome this challenge, there's been a growing demand in developing novel dielectric, medium-refractive indexed nanoparticles. These materials can support light manipulation at the nanoscale and still exhibit strong extinction cross-sections.[16] The dielectric nanostructures exhibit both the electric and magnetic multipolar modes that can lead to enhanced optical responses, such as nanoantennas, optical sensors, surface-enhanced absorption, fluorescence, and Raman scattering.[16] While the PMNs, can show only electric multipoles (dipoles and quadrupoles) exhibiting limitations.

Mohammadparast et al. have recently shown how strongly Mie resonances in dielectric $Cu_2O$ cubes can be tailored through size. Through finite-difference time-domain (FDTD) simulations, the authors have shown that $Cu_2O$ cubes exhibit strong enhancement of electric as well as magnetic fields at the Mie resonances.[17] Some of the notable findings include tunability of Mie resonance in $Cu_2O$ using different shapes and sizes indicating evidence of strong scattering cross section from simulations and experiments, and high scattering cross section of cubes over spheres for similar size. All this is possible in $Cu_2O$ due to its inherent property of being dielectric, moderate refractive indexes (RI) material having a value of $n = 2.2620$ & $k = 0$ at bandgap. Also, the work compares plasmonic silver (Ag) and $Cu_2O$ through simulations on the field enhancement to understand how these materials respond under light interaction. The range of moderate RI (1.7 – 3.0). It is critical to have materials in this range because for dielectric materials with RI greater than 3, the magnetic and electric resonances are spectrally separated from each other, which inhibits their use in directional scattering at their total scattering peaks. The findings of this work are an extension of the work done by Mohammadparast et al. as the basis, we extend the scope and concept to another metal-oxide known as cupric oxide (CuO), whose RI (n = 2.1~2.37, k = 0) at the bandgap. In this work, the focus shall remain pertinent to CuO synthesis, characterization, simulation explaining the Mie resonances contribution from both electric and magnetic fields in the visible to near-infrared region.

RESULTS AND DISCUSSION:

To understand the effect of Mie resonances in CuO particles synthesized CuO nanoparticles primarily by oxidation from $Cu_2O$ cubic nanoparticles in the tubular furnace at 500°C for three hours by flowing air. The detailed procedure for making $Cu_2O$ and CuO nanoparticles are mentioned in the supporting information (SI). The schematic diagram for $Cu_2O$ to CuO is shown in Scheme 1.

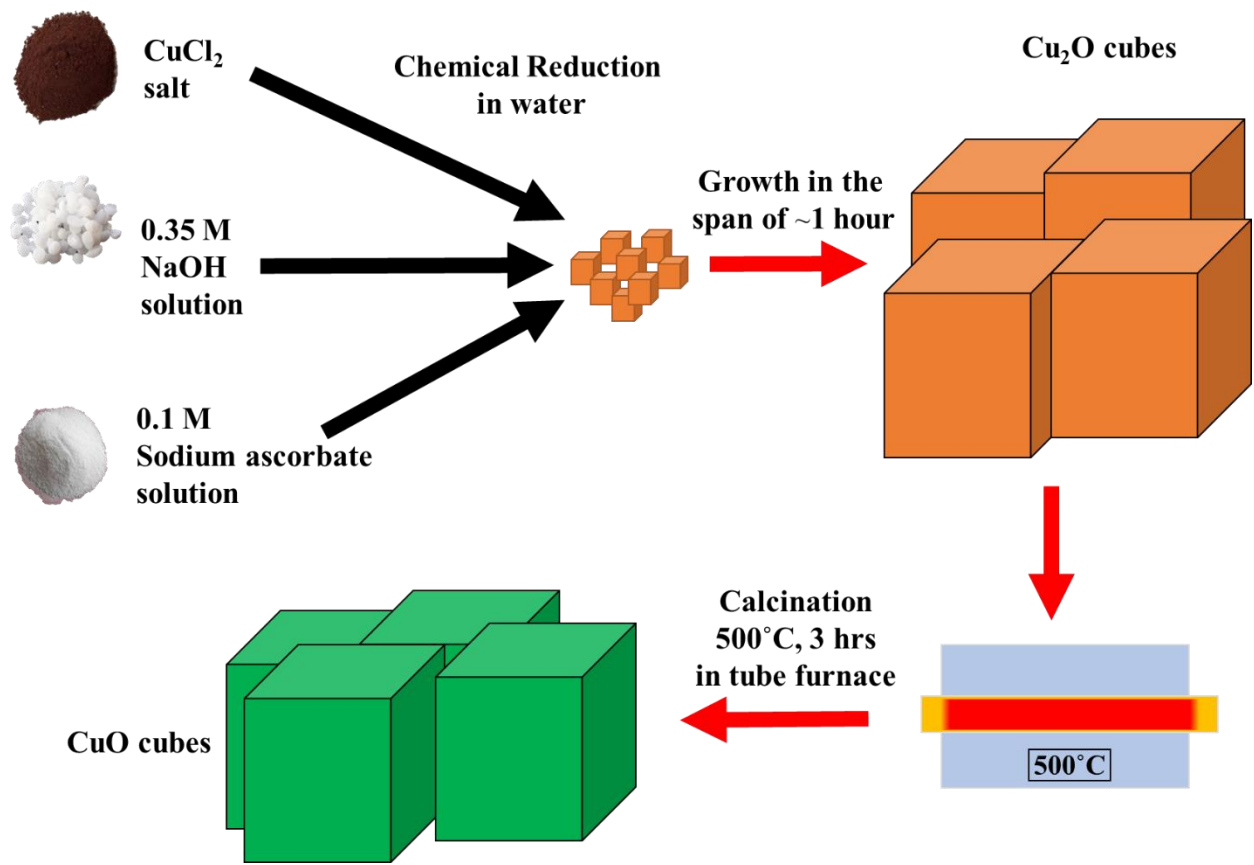

**Scheme 1**: Synthesis of CuO cubes through oxidation of $Cu_2O$ cubes performed at 500°C for 3 hr. in continuous supply of air.

The X-ray diffraction (XRD) shows $Cu_2O$ (Figure 2a) and CuO (Figure 2b). The corresponding TEM images are shown in Figures 2c and 2d. The particle size measured before oxidation is 426± 63 nm. From Figure 2d we see that the morphology of CuO changes slightly from cubic ($Cu_2O$) to spherical after oxidation, this change is attributed to the aggregation of particles at high temperature. We may note that this morphology change results in the change in density, for CuO (6.31 g/cm$^3$) which is slightly higher compared to $Cu_2O$ (6 g/cm$^3$). $Cu_2O$ is a p-type semiconductor with a bandgap of ~ 2.1 eV. The diffuse-reflectance spectroscopy (DR-UV-Vis) of before and after oxidation are shown in Figure 2e-f. In addition, we performed FDTD simulations for different particles sizes and shapes (Figure 3). The detailed explanation for the FDTD simulation procedure can be seen in SI. The simulated extinction spectra for the CuO sphere and cubes are shown in Figure 3a-b. The Mie resonance peaks of CuO spheres ranging 25-175 nm (Figure 3a) show the resonance peak keeps increasing as particle size increases and also red-shifts. The peak maximum shifts the simulation window of 350-2000 nm for particle size of 125 nm and larger. When the particle size increases to 200 nm and above (Figure 3b), there are multiple Mie resonance peaks spread across the simulation window. The largest spherical particle is 400 nm in diameter, two of the three resonance peaks are below the bandgap of CuO (i.e., wavelength above 1033 nm). CuO is a p-type semiconductor with bandgaps of ~1.2 eV.[18] These resonance features are not seen in bulk single crystalline CuO. The trend is similar for CuO cubic particles, as well. With increasing particle size, the resonance peak energy redshifts (moves lower energy). The simulated scattering (Figure S1a and S2a) and absorption spectra (Figure S1b and S2b) reveal that scattering is the dominant contribution to Mie extinction for the investigated particle sizes (both sphere and cubes). Hence, this sums up the fact that by varying shape and size we can control the fraction light in the UV-Vis-NIR-light region, with scattering as the dominant feature. This also underscores the point that these CuO can be used as strong scattering material by directing light in the forward direction.

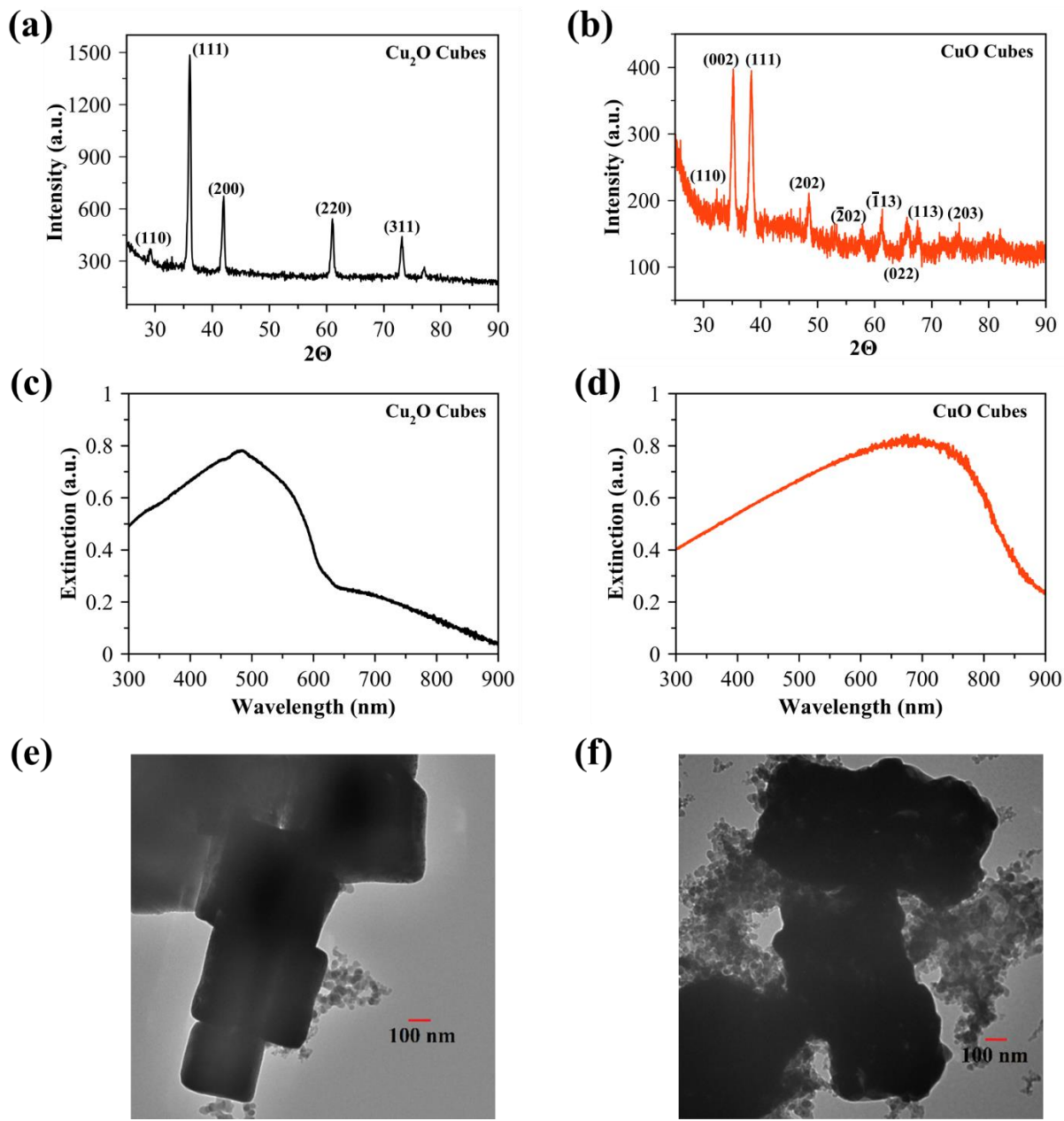

**Figure 2.** The representative **(a, b)** X-ray diffraction pattern of $Cu_2O$ and $CuO$ cubes. **(c, d)** Diffusive reflectance UV-Vis spectra of $Cu_2O$ and $CuO$ cubes. **(e, f)** Representative TEM images of $Cu_2O$ (426± 63 nm) and CuO cubes (530± 49 nm).

To have a better understanding of the Mie resonance peaks observed in the extinction spectra of CuO particles, we performed a detailed analysis on the simulation data for 175 nm spherical particles, which we took as the basis. The simulated extinction spectrum of the 175 nm diameter CuO sphere (Figure 4a) and a comparative study with a plasmonic material of 175 nm diameter Ag sphere (Figure 4b). The CuO particle exhibits a single resonance peak, at 514 nm. The magnetic and electric field spatial distributions at the resonance peak respectively are shown in Figure 4c-d. Similarly, the field distributions for the plasmonic material are seen in Figure 4e-f. For generating the field distributions in the simulation, a plane wave is used for propagation in the direction of the positive x-axis and polarized along the y- and the z-axes for the electric and magnetic fields, respectively. The detailed simulation procedure is described in the SI.

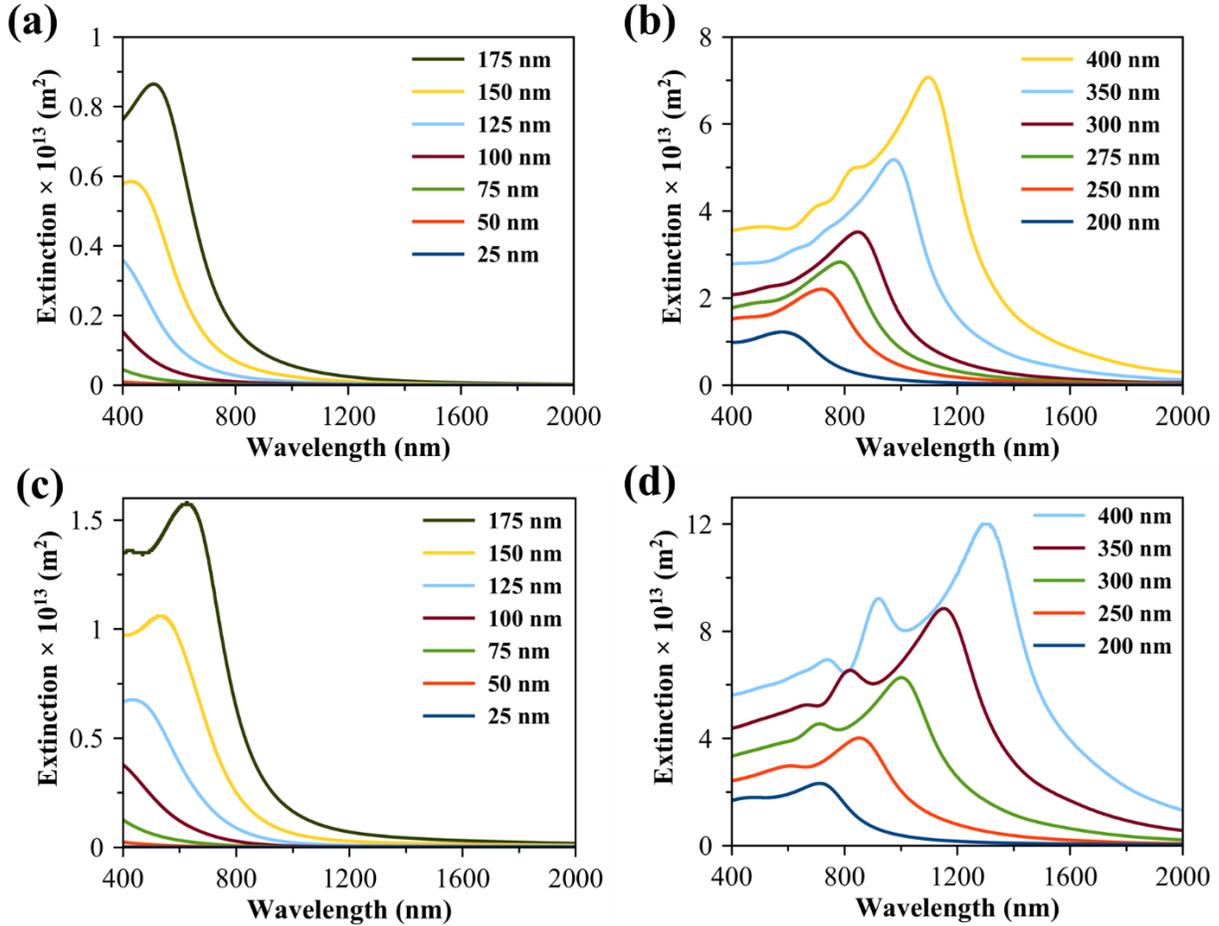

**Figure 3.** Simulated extinction cross section as a function of wavelength for CuO spheres of various diameters ranging from **(a)** 25 to 175 nm, and **(b)** 200 to 400 nm. Simulated extinction cross section as a function of wavelength for CuO cubes of various edge lengths ranging from **(c)** 25 to 175 nm, and **(d)** 200 to 400 nm. The source illumination direction is perpendicular to the principal axis.

The electric field patterns (Figure 4d) show a two-lobe distribution, which is a characteristic trait of an electric dipole. Similarly, the magnetic field patterns (Figure 4c) confirm the presence of magnetic dipole mode.[19] The enhanced magnetic field patterns (Figure 4c) are generated by the circular displacement currents, resulting from a strongly coupled electric field penetrating the particle.[16] The resulting magnetic response is made possible when the particle size (diameter, d) and wavelength of the incident light (λ) penetrating it are comparable (d ~ λ/n), n is the refractive index.[16] Thus explaining the tunability of the resonance with geometry.

The enhancements of magnetic and electric fields up to 5 and 6 times, respectively, over the incident far-field (Figures 4c-d). On comparing the magnetic and electric field distributions (Figure 4c-d and S5a-b) near 514 nm, almost equal magnitude in field strength. The field distributions near 514 nm show dipole enhancement in both electric and magnetic. On comparison with PMN, we provide the simulated results of the Ag sphere at the LSPR peak wavelength occurring at 530 nm (Figures 4e-f and S6a-b). Results show that the magnetic field enhancement is relatively weak compared to the electric for the Ag particle. Hence, the electric dipole mode dominates the LSPR at 530 nm.

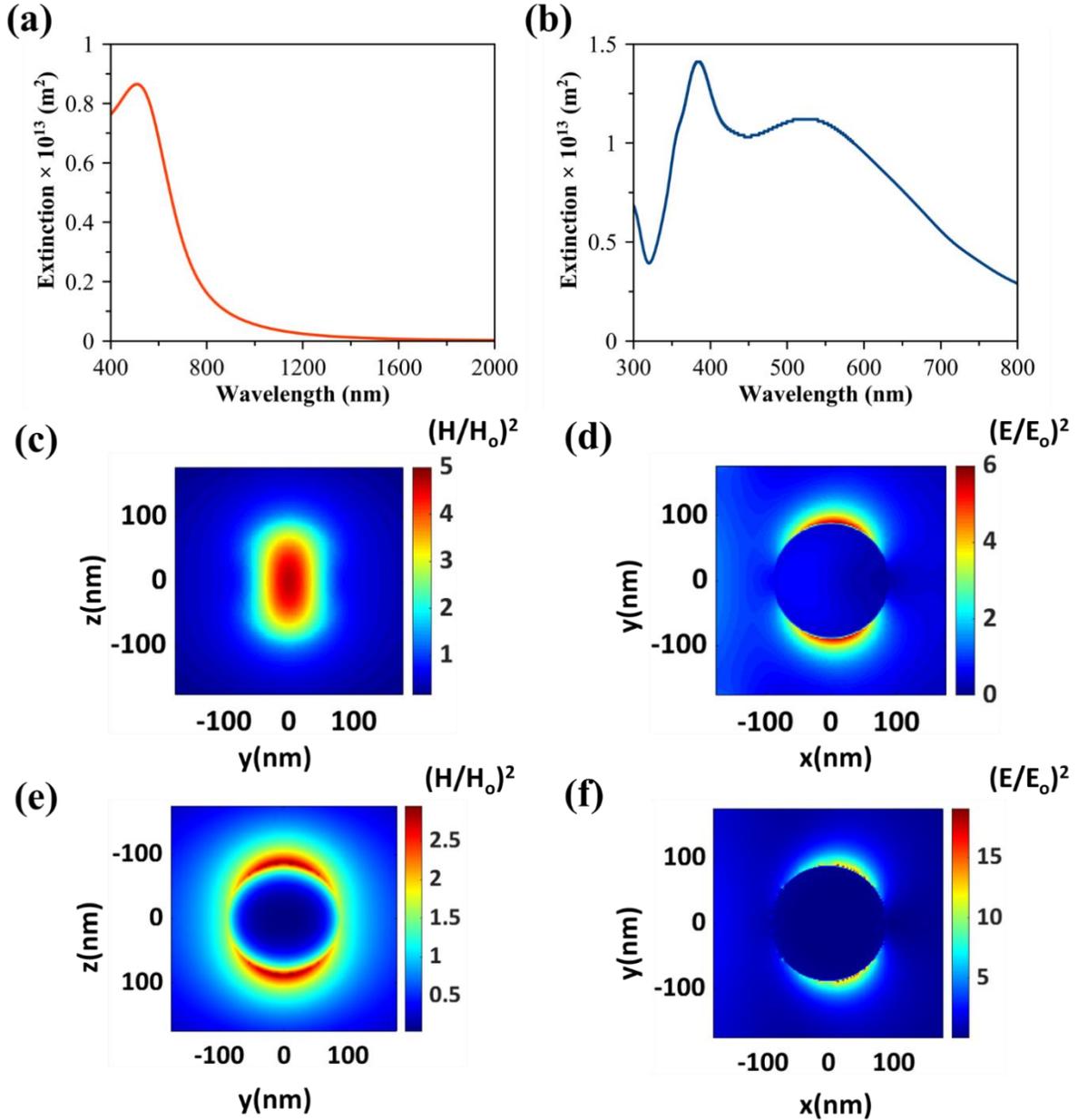

**Figure 4.** Simulated extinction cross section as a function of wavelength for (**a**) 175 nm CuO sphere, and (**b**) 175 nm Ag sphere. Simulated spatial distribution of enhancement in (**c**) magnetic field intensity [$H^2/H_0^2$] in YZ plane, and (**d**) electric field intensity [$E^2/E_0^2$] in XY plane, for 175 nm CuO sphere at the resonance wavelength of 514 nm. Simulated spatial distribution of enhancement in (**e**) magnetic field intensity [$H^2/H_0^2$] in YZ plane, and (**f**) electric field intensity [$E^2/E_0^2$] in XY plane, for 175 nm Ag sphere at the resonance wavelength of 530 nm.

While analyzing the multiple Mie resonance peaks observed in the extinction spectrum of larger CuO particles, in this case 400 nm cubic particles (Figure 3d), we analyzed the magnetic and electric field distributions at multiple wavelengths around the lowest and second lowest energy resonance peaks. The magnetic and electric field distributions at the resonance peak (Figure 5a-b), respectively, for the 400 nm CuO cubic particle at the lowest energy (1296 nm) and second lowest energy (935 nm). The results also show that the 400 nm CuO cubic particle can exhibit localization and enhancement of up to 80 and 15 orders of magnitude for the magnetic and electric near field intensities, respectively, over the incident far field. From the FDTD simulation results (Figure 5a-b) along with the results shown in Figures S7a-b, it can be concluded that the lowest energy Mie resonance peak is due to the combination of the magnetic and electric dipole excitations. To elucidate the second-lowest energy Mie resonance peak (935 nm) observed in the extinction spectrum (Figure 5c-d), we present the magnetic and electric field distributions at 935 nm. From Figures 5c-d and S7c-d it is concluded that both the magnetic and electric field quadrupole excitations contribute to this resonance peak.

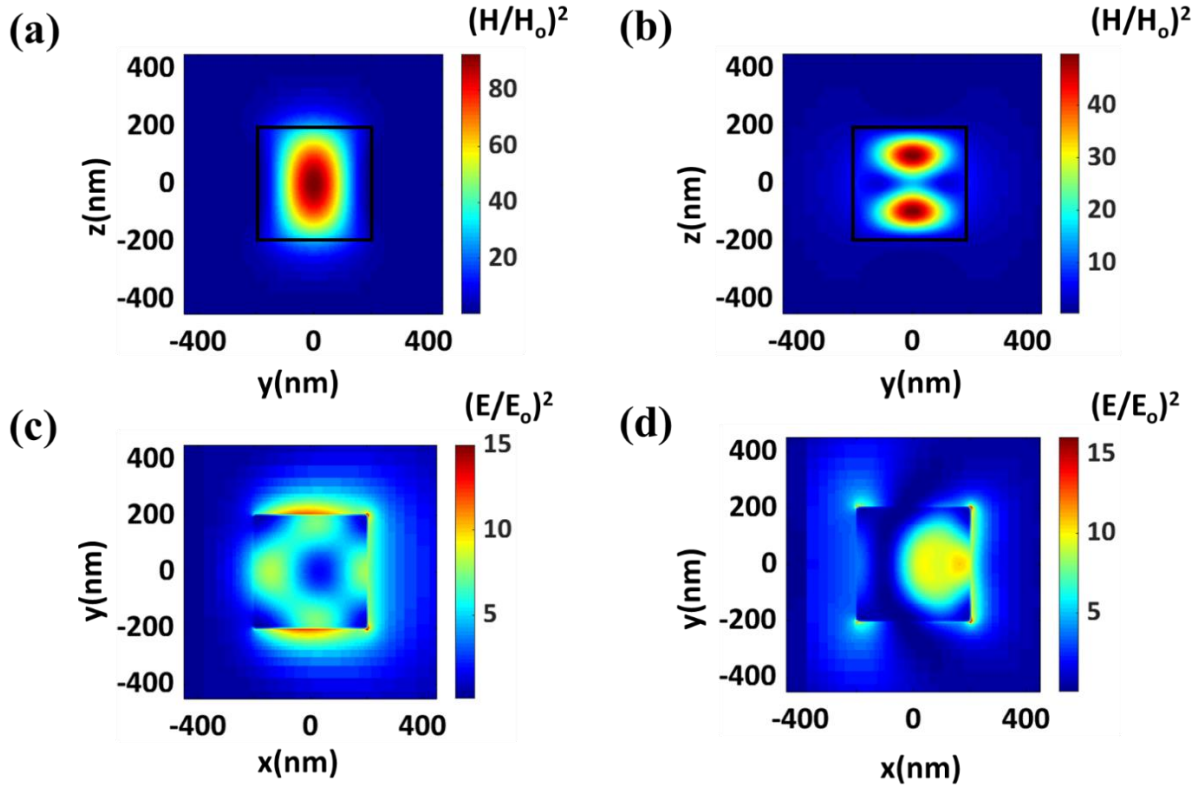

**Figure 5.** Simulated spatial distribution of enhancement in **(a and c)** magnetic field intensity $[H^2/H_0^2]$ in YZ plane, and **(b and d)** electric field intensity $[E^2/E_0^2]$ in XY plane for CuO cubes of 400 nm edge length at the resonance wavelength of **(a and b)** 1296 nm and **(c and d)** 935 nm.

## CONCLUSION

In conclusion, we present this work as an extension of the previous work done by our group on Mie resonance on $Cu_2O$ cubic particles. Here, we demonstrate through experimental and simulation that Mie resonances in subwavelength CuO cubes particles are tunable with geometry. Even the lowest-energy Mie resonance extinction spectrum observed is an energetic combination of the magnetic and electric dipole excitations. Similarly, for higher energy regions we observe contribution from both the magnetic and electric quadrupole excitations in the extinction spectrum. The tunability of the Mie resonance peaks in the visible-to-near-IR region can be achieved through controlling geometry. The subwavelength CuO cubes show a higher scattering cross section compared to spherical particles of similar size. In addition, being a strong dielectric resonator these materials strong exhibit electric and magnetic fields enhancement for the incident light at Mie resonances which are comparable or slightly larger than plasmonic Ag particles. All these properties make CuO to be an effective material as nanoantenna for concentrating and directing light towards facilitating photocatalysis. In addition, they are a good alternative to PMNs for other applications but not limited to nanophotonic, solar-light harvesting, and spectroscopy.

**Supporting Information:**

*Sundaram Bhardwaj Ramakrishnan[⊥], Ravi Teja Addanki Tirumala[⊥], Farshid Mohammadparast[⊥], Swetha M. Arumugam[§], Marimuthu Andiappan*[⊥]*

Affiliations:

[⊥] School of Chemical Engineering, Oklahoma State University, Stillwater, OK, USA.

[§] Department of Chemistry, PSG College of Arts & Science, Coimbatore, Tamil Nadu, India.

## I. Preparation of Cu$_2$O particles

**Microemulsion method for Cu$_2$O spheres:** The spherical Cu$_2$O particles were prepared using the water-in-oil microemulsion synthesis method. In this method, n-heptane, polyethylene glycol-dodecyl ether (Brij, average Mn ~362), copper nitrate, and hydrazine were used as the continuous oil phase, surfactant, copper precursor, and reducing agent, respectively. These chemicals were added in the following sequence and quantity. First, 7.75 mL of n-heptane was added to the three-neck round bottom flask at room temperature, followed by 1.15 mL of Brij surfactant. During the process of surfactant addition, the system was stirred vigorously, then 0.81 mL of 0.1 M copper nitrate aqueous solution was added, followed by 0.81 mL of 1 M of aqueous hydrazine solution. To understand the evolution of extinction spectra as a function of particle size, samples were taken at different time intervals for UV-Vis-near IR extinction measurements.

**Chemical reduction method for Cu$_2$O cubes:** The chemical reduction method reported in the literature was used for the synthesis of Cu$_2$O cubes.[1] Using this method, we first made 30 mL of 0.0032 M aqueous solution of CuCl$_2$, which acts as a copper source. This quantity of 30 mL is poured into a three-neck round bottom flask, which is placed under an inert atmosphere filled with nitrogen. To this solution at room temperature, we added 1 mL of 0.35 M aqueous NaOH solution, which is expected to result in Cu (OH)$_2$ colloids formation immediately.[1] Then, 1 mL of sodium ascorbate (reducing agent) was added. The solution then became brick-red, indicating the formation of Cu$_2$O particles[1]. The synthesis was done at room temperature and nitrogen environment.

## II Preparation of CuO particles

The CuO particles were prepared through the oxidation of Cu$_2$O particles placed in vials inside the furnace through a glass reactor. At the start, the atmosphere inside the reactor was passed through air to keep the system in the desired condition. Followed by which temperature of the reactor was kept at 500°C for 3 hours with a continuous supply of air for complete oxidation of Cu$_2$O.

**III. Characterization of Cu$_2$O particles**

**Transmission Electron Microscope (TEM) Imaging:** The TEM images were taken using JEOL JEM-2100 TEM and Thermo Fisher Scientific Titan Themis 200 G2 aberration-corrected TEM. The JEOL JEM-2100 system is equipped with a LaB6 gun and an accelerating voltage of 200 kV. The Titan Themis 200 system is equipped with a Schottky field-emission electron gun and operated at 200 kV. Samples were prepared by taking 150 µL of the sample into 2 mL of deionized water. This dilution solution was sonicated to ensure proper dispersion. Next, 10 µL was taken and placed onto a TEM grid with ultra-thin carbon supporting the film. The water was allowed to evaporate at room temperature before TEM imaging.

**UV-Vis-near IR extinction measurements:** All UV-Vis-near IR extinction spectra were taken using an Agilent Cary 60 Spectrophotometer. 150 µL of the synthesis mixture was diluted into 2 mL of deionized water. This diluted sample was then used for UV-Vis-near IR extinction measurements.

**III. Details of finite-difference time-domain (FDTD) Simulations**

To implement FDTD simulations, we employed the Lumerical FDTD package.[2] The optical properties of Ag, Cu$_2$O, and CuO were taken from Palik.[3] The real (n) and imaginary (k) parts of the refractive index used in the simulations for Cu$_2$O and CuO are shown in Table S1. The boundary conditions used for the simulations were perfectly matched layer (PML) conditions in the x, y, and z directions. For the simulations of extinction, scattering, and absorption spectra, the respective cross sections as a function of wavelengths were calculated using the total-field/scattered-field (TFSF) formalism. The incident light source used for these simulations was the Gaussian source in the simulated wavelength region. For the simulations of the magnetic and electric field distributions, a plane wave was used for electromagnetic field incidence with propagation in the x-axis direction, and polarization along the y-axis and the z-axis for the electric field and the magnetic field, respectively. For all the simulation results are shown in the main draft, and the directly associated supporting figures (i.e., Figures S3a-d, S4a-d, S5a-d, S6a-d, S7-S14, and S15a-d)), the simulations were performed for Cu$_2$O or CuO particles in a surrounding medium with the refractive index of 1. We also simulated the effect of the refractive index of the medium on the Mie resonance of Cu$_2$O and CuO particles. For these simulations, we used the refractive index of water (~1.33). Figure 16 and Figure 17 show the representative simulation results with a medium refractive index of 1.33. By the comparison of Figure S4a-d (i.e., a refractive index of 1) and Figure S17a-d (i.e., a refractive index of 1.33), the magnetic field and electric field enhancement values are damped for the refractive index of 1.33.

**CuO Spheres:**

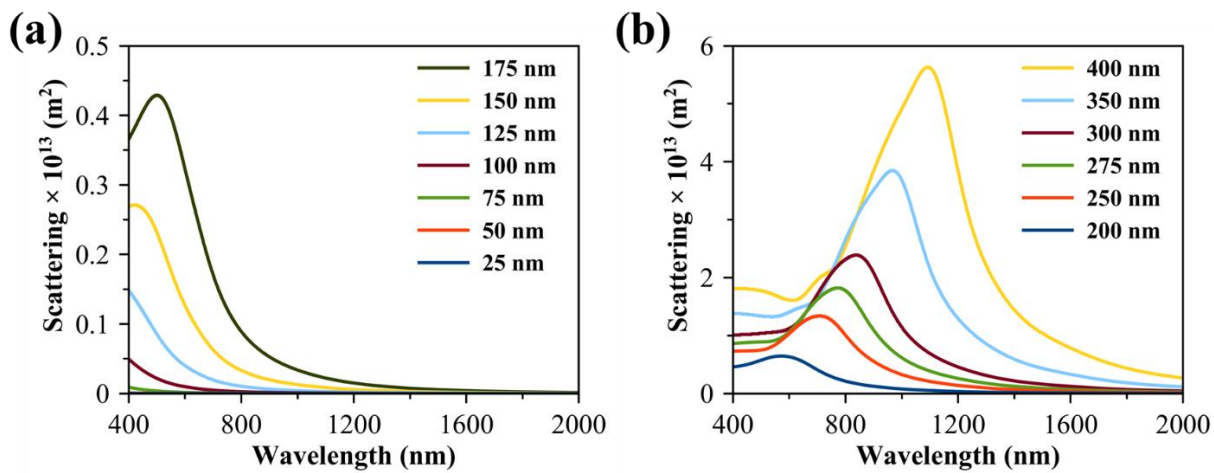

**Figure S1.** Simulated scattering cross section as a function of wavelength for CuO spheres of various diameters ranging from **(a)** 25 to 175 nm, and **(b)** 200 to 400 nm.

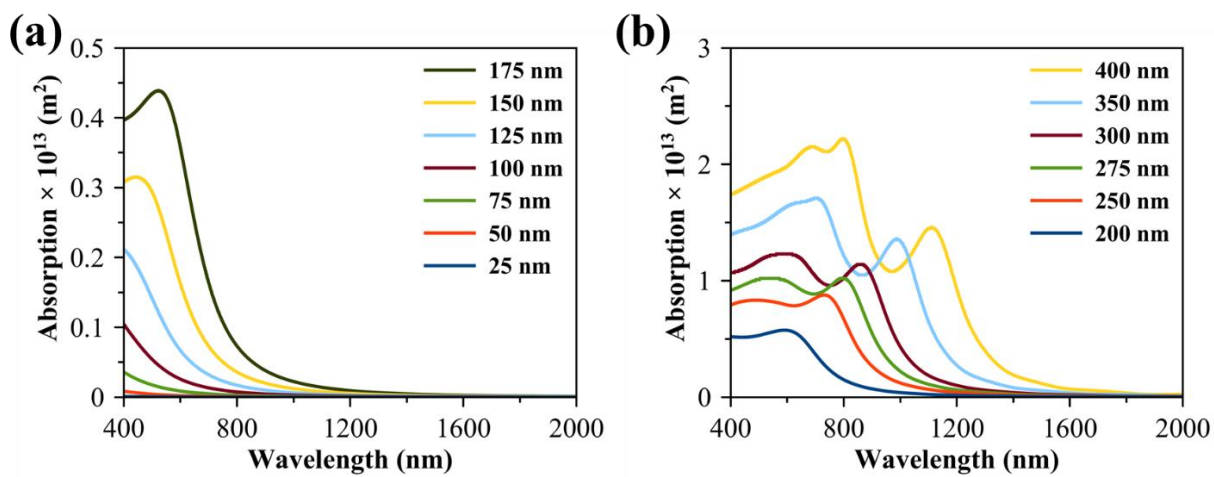

**Figure S2.** Simulated absorption cross section as a function of wavelength for CuO spheres of various diameters ranging from **(a)** 25 to 175 nm, and **(b)** 200 to 400 nm.

**CuO Cubes:**

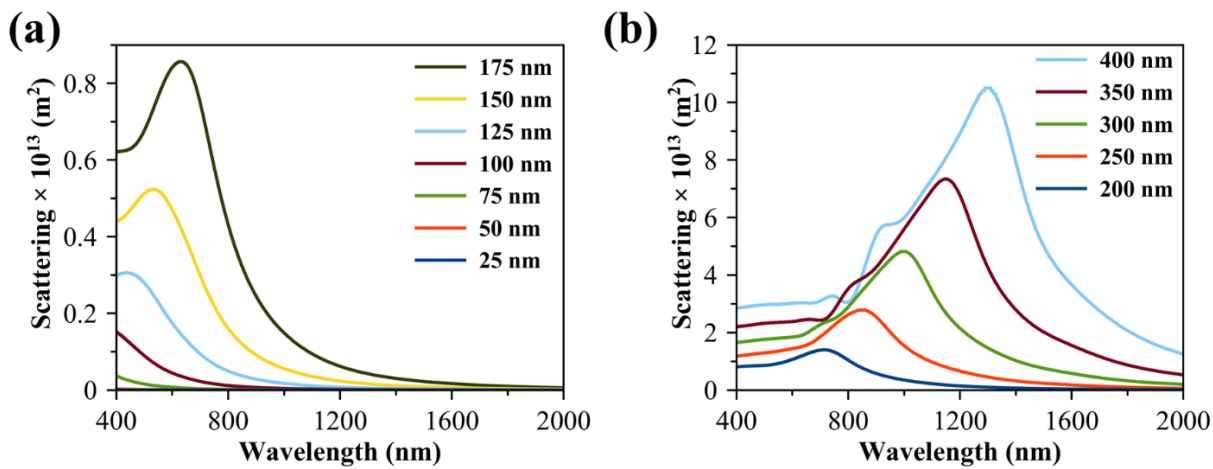

**Figure S3.** Simulated scattering cross section as a function of wavelength for CuO cubes of various edge lengths ranging from **(a)** 25 to 175 nm, and **(b)** 200 to 400 nm. The source illumination direction is perpendicular to the principal axis.

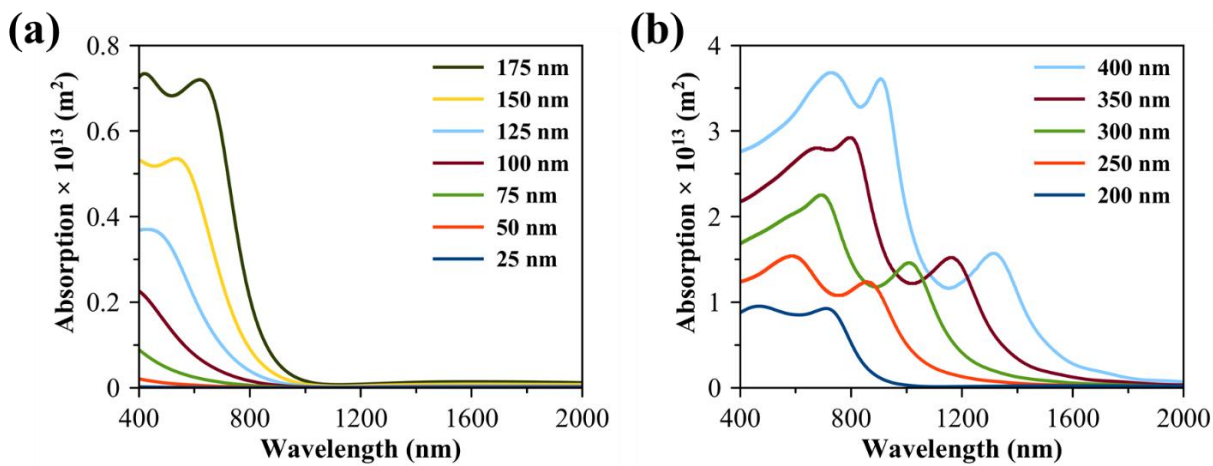

**Figure S4.** Simulated absorption cross section as a function of wavelength for CuO cubes of various edge lengths ranging from **(a)** 25 to 175 nm, and **(b)** 200 to 400 nm. The source illumination direction is perpendicular to the principal axis.

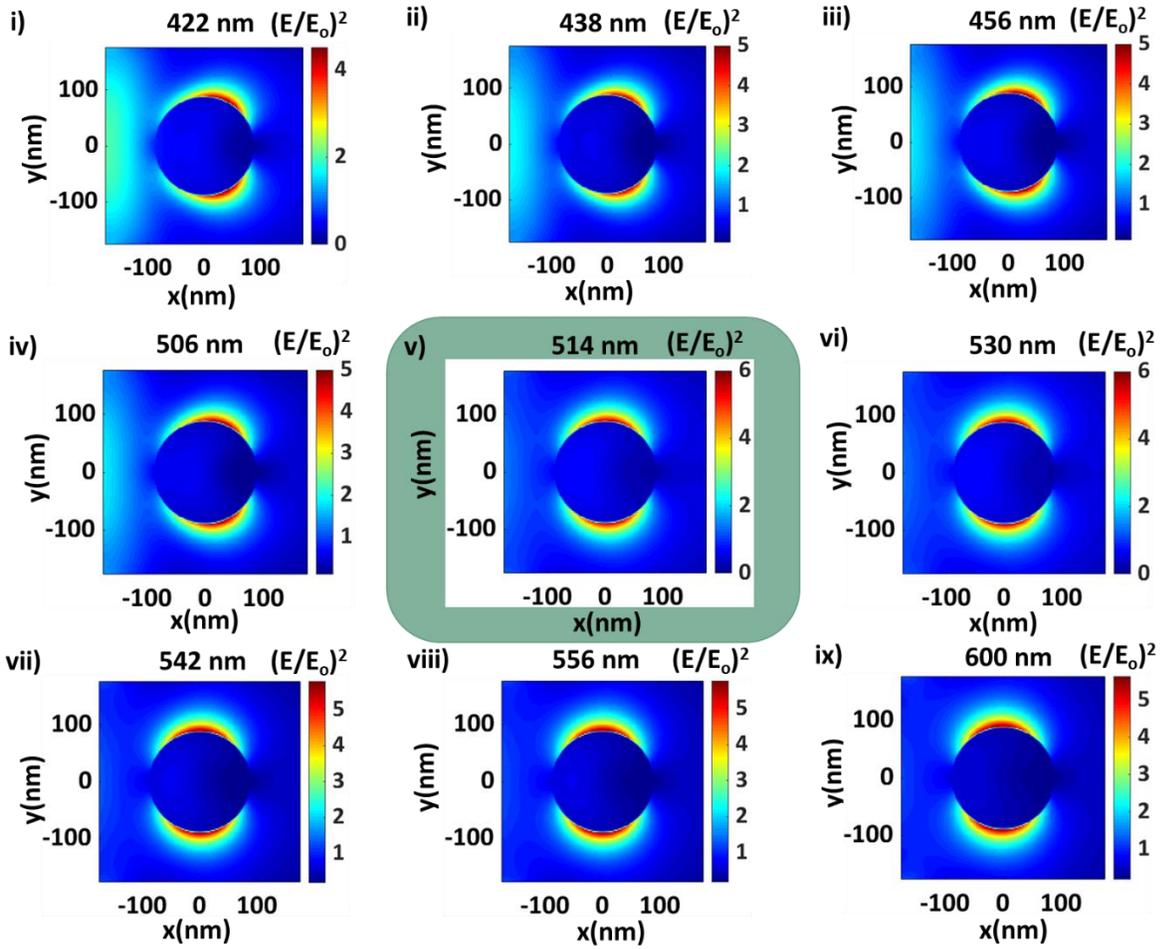

**Figure S5a.** Simulated spatial distribution of enhancement in electric field intensity [$E^2/E_0^2$] in XY plane at different wavelengths across the Mie resonance peak wavelength (i.e., **514 nm**) for CuO sphere of 175 nm diameter.

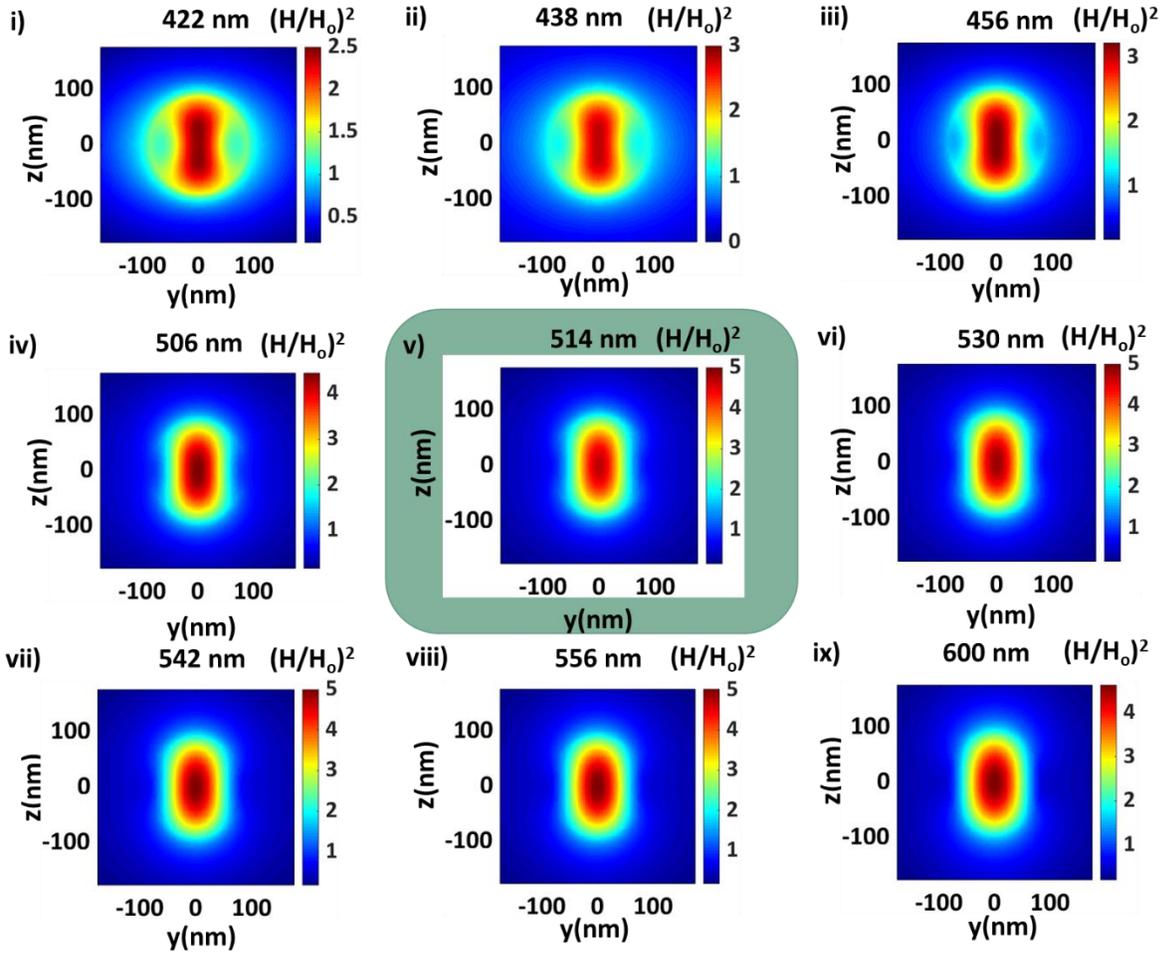

**Figure S5b.** Simulated spatial distribution of enhancement in magnetic field intensity [$H^2/H_0^2$] in XY plane at different wavelengths across the Mie resonance peak wavelength (i.e., **514 nm**) for CuO sphere of 175 nm edge length.

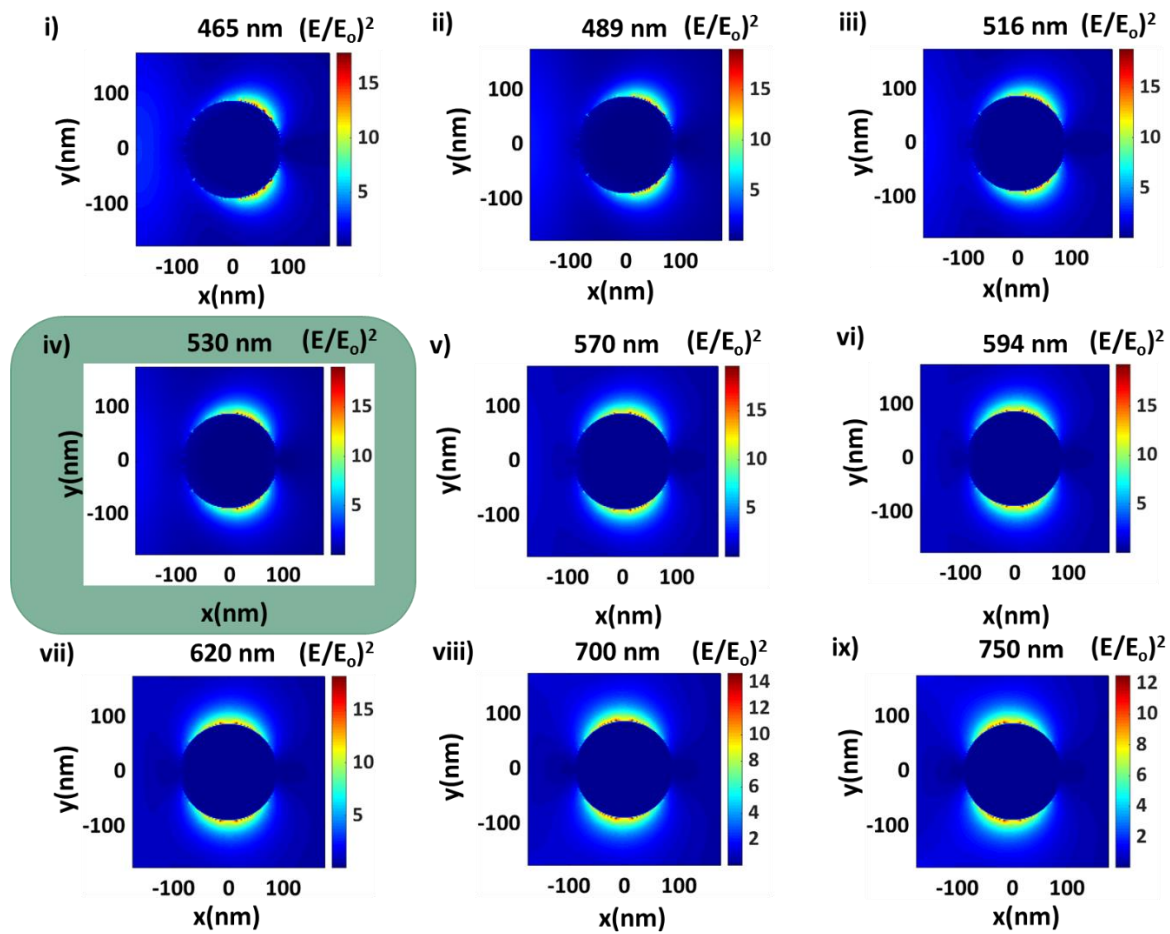

**Figure S6a.** Simulated spatial distribution of enhancement in electric field intensity [$E^2/E_0^2$] in XY plane at different wavelengths across the Mie resonance peak wavelength (i.e., **530 nm**) for Ag sphere of 175 nm diameter.

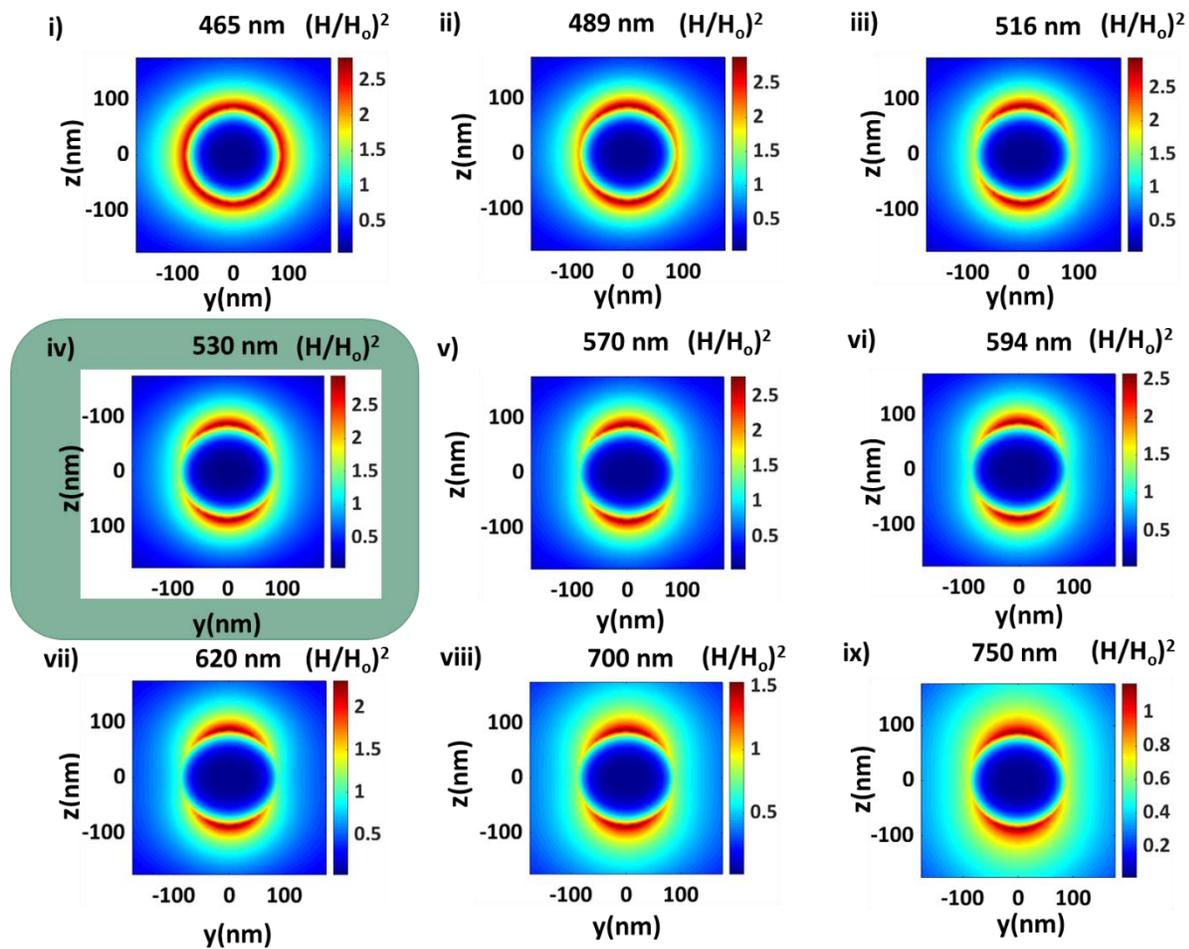

**Figure S6b.** Simulated spatial distribution of enhancement in magnetic field intensity [$H^2/H_0^2$] in XY plane at different wavelengths across the Mie resonance peak wavelength (i.e., **530 nm**) for Ag sphere of 175 nm diameter.

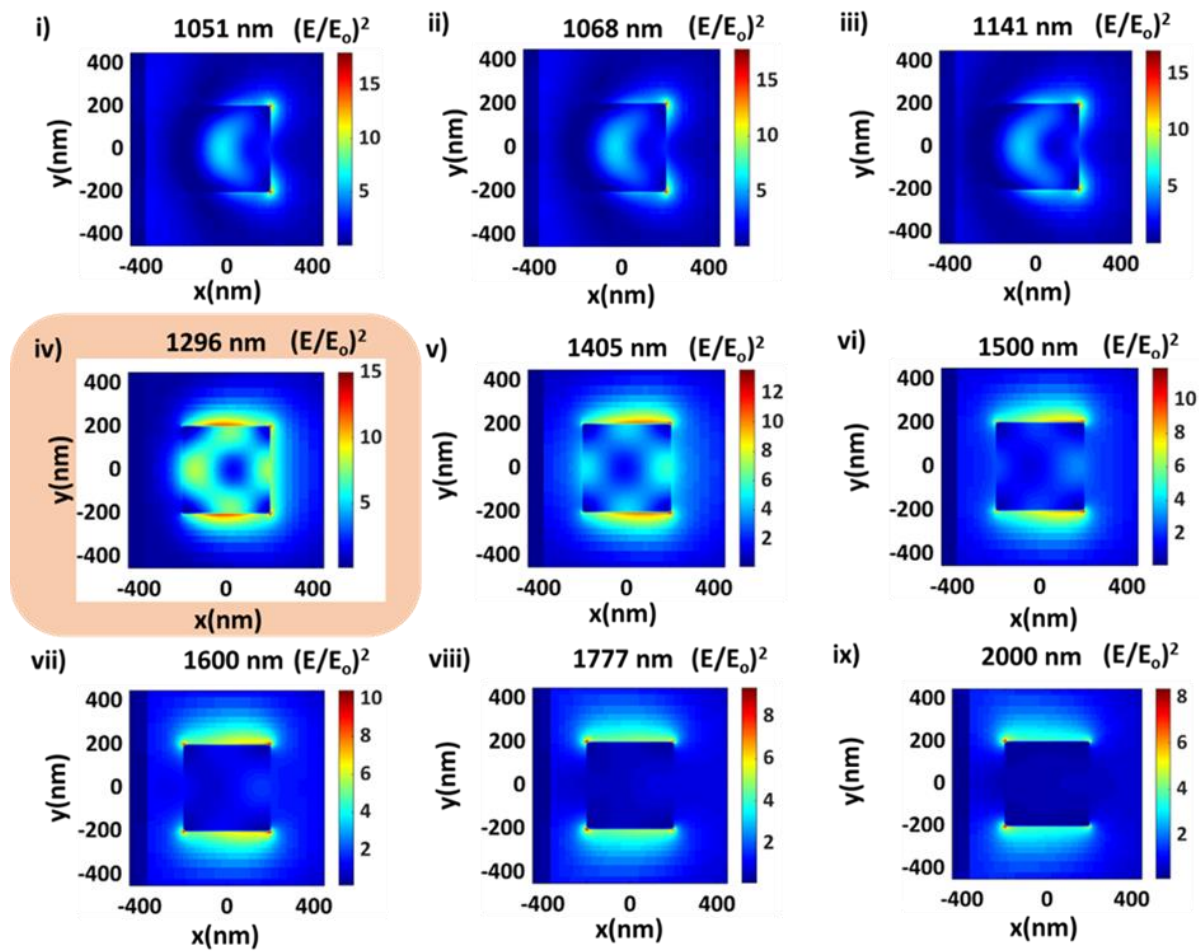

**Figure S7a.** Simulated spatial distribution of enhancement in electric field intensity [$E^2/E_0^2$] in XY plane at different wavelengths across the lowest energy Mie resonance peak wavelength (i.e., **1296 nm**) for CuO cube of 400 nm edge length.

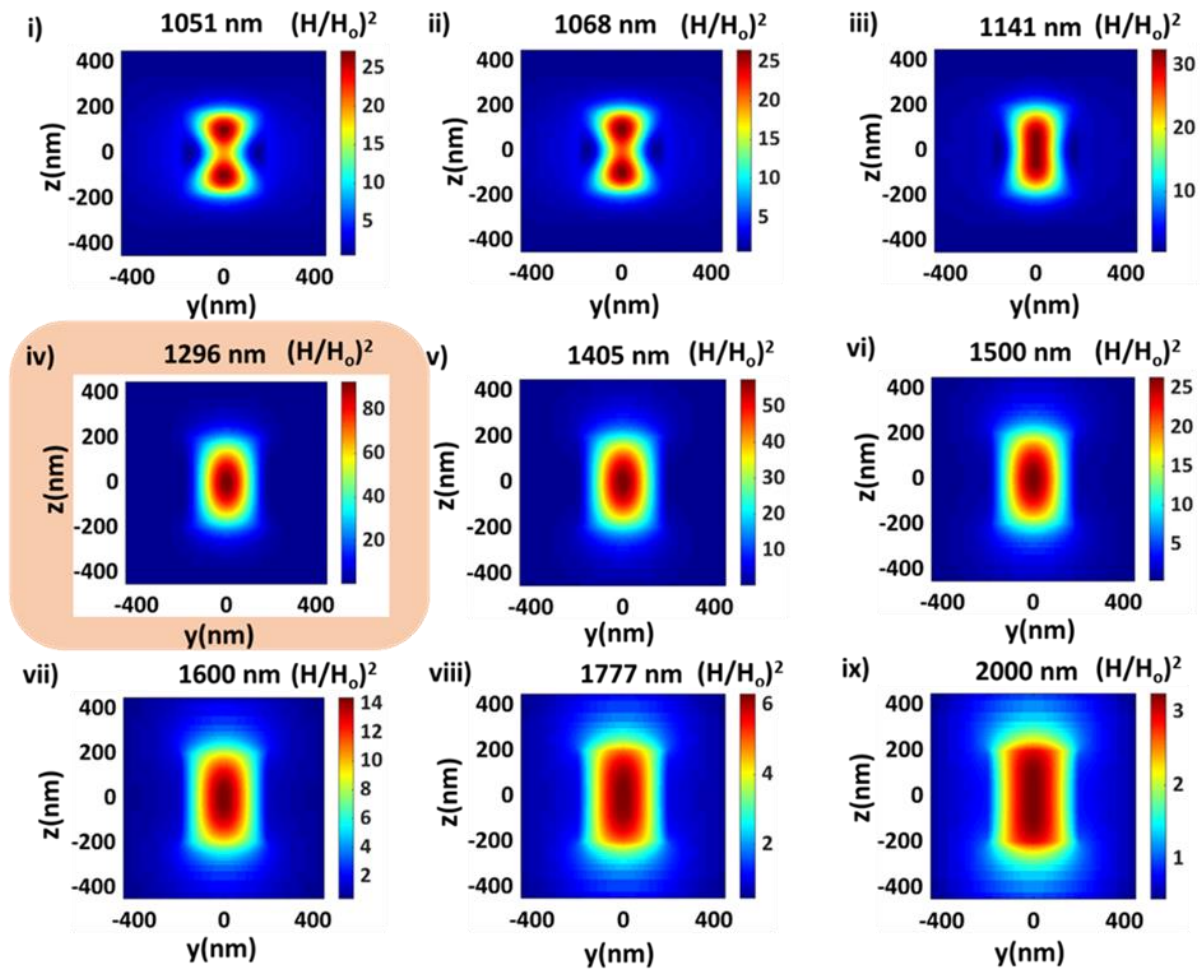

**Figure S7b.** Simulated spatial distribution of enhancement in magnetic field intensity [$H^2/H_0^2$] in YZ plane at different wavelengths across the lowest energy Mie resonance peak wavelength (i.e., **1296 nm**) for CuO cube of 400 nm edge length.

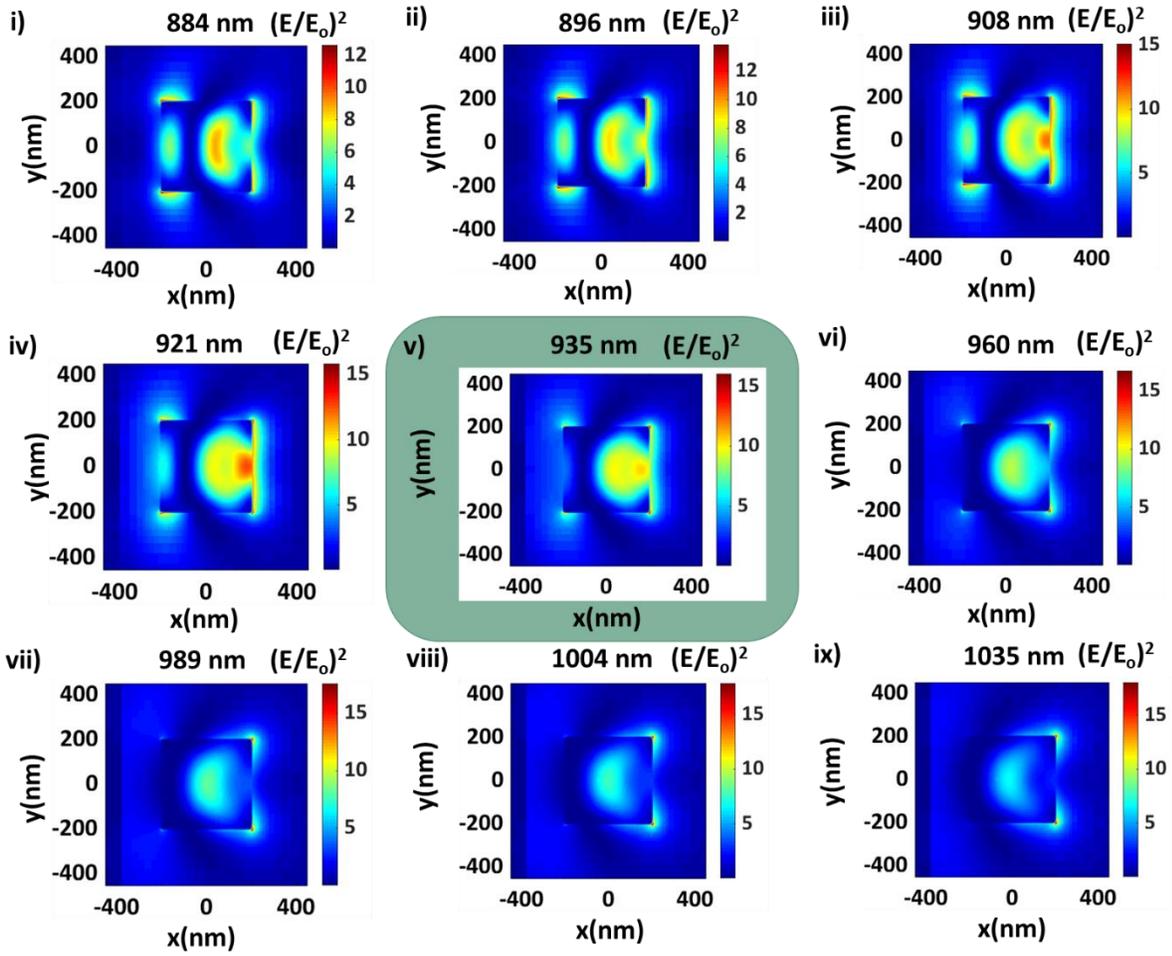

**Figure S7c.** Simulated spatial distribution of enhancement in electric field intensity [$E^2/E_0^2$] in XY plane at different wavelengths across the second-lowest energy Mie resonance peak wavelength (i.e., **935 nm**) for CuO cube of 400 nm edge length.

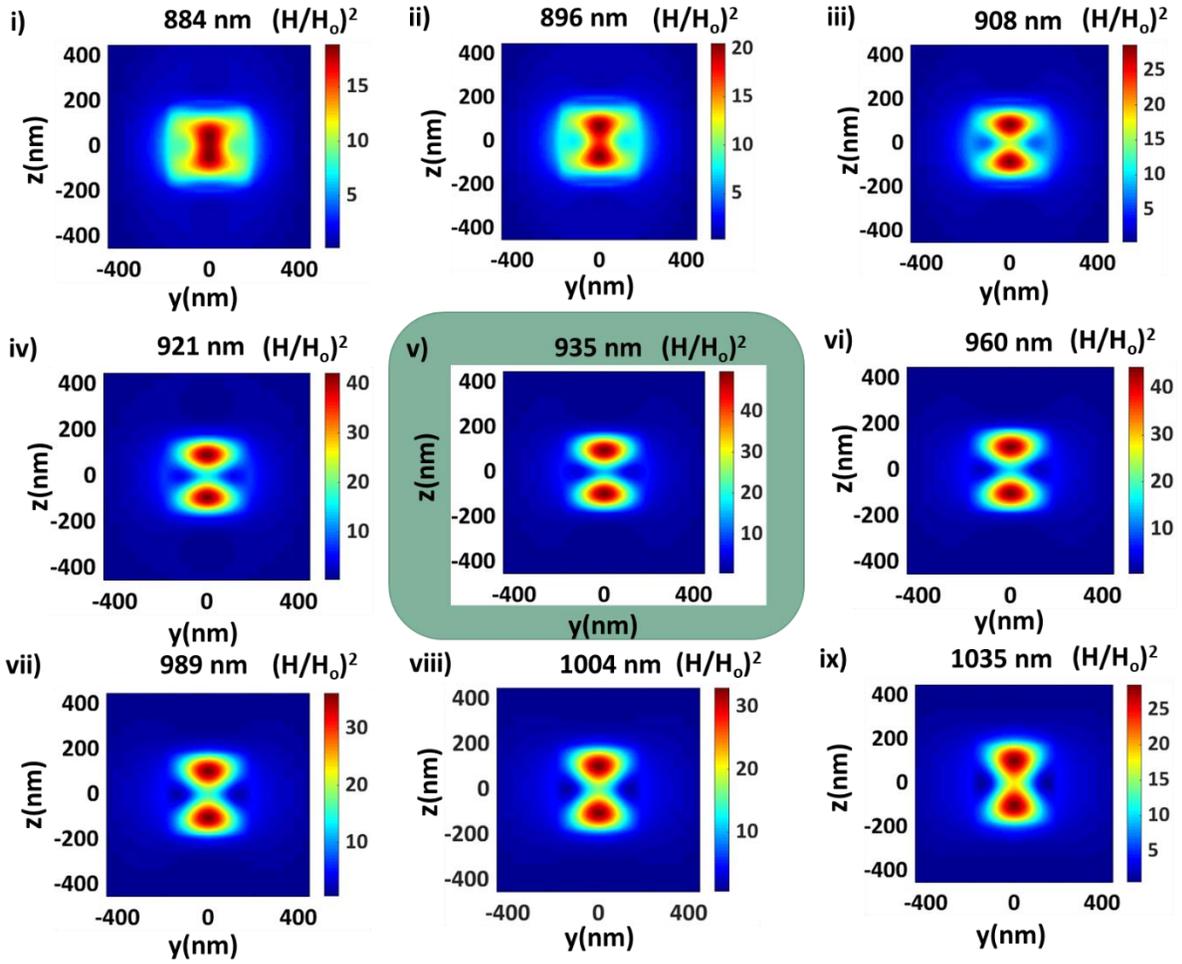

**Figure S7d.** Simulated spatial distribution of enhancement in magnetic field intensity [$H^2/H_0^2$] in YZ plane at different wavelengths across the second-lowest energy Mie resonance peak wavelength (i.e., **935 nm**) for CuO cube of 400 nm edge length.